\documentclass[aps,prl,showpacs,twocolumn,preprintnumbers,amsmath,amssymb]{revtex4}

\usepackage{graphicx}% Include figure files

\begin{document}

\title{Strong coupling in multimode quantum electromechanics}% Force line breaks with \\

\author{Atsushi~Noguchi$^{1}$}
\email[]{noguchi@qc.rcast.u-tokyo.ac.jp}
\author{Rekishu~Yamazaki$^{1,2}$}
\author{Manabu~Ataka$^{3}$}
\author{Hiroyuki~Fujita$^{3}$}
\author{Yutaka~Tabuchi$^{1}$}
\author{Toyofumi~Ishikawa$^{1}$}
\author{Koji~Usami$^{1}$}
\author{Yasunobu~Nakamura$^{1,4}$}

\affiliation{%
$^{1}$Research Center for Advanced Science and Technology (RCAST), The University of Tokyo, Meguro-ku, Tokyo, 153-8904, Japan
}
\affiliation{%
$^{2}$PRESTO, Japan Science and Technology Agency, Kawaguchi-shi, Saitama 332-0012, Japan
}
\affiliation{%
$^{3}$Institute of Industrial Science, The University of Tokyo, Meguro-ku, Tokyo, 153-8505, Japan
}
\affiliation{%
$^{4}$Center for Emergnent Matter Science (CEMS), RIKEN, Wako-shi, Saitama 351-0198, Japan
}

\date{\today}% It is always \today, today,
             %  but any date may be explicitly specified
             
\begin{abstract}
Cavity electro-(opto-)mechanics allows us to access not only single isolated but also multiple mechanical modes in a massive object. 
Here we develop a multi-mode electromechanical system in which a several membrane vibrational modes are coupled to a three-dimensional loop-gap superconducting microwave cavity.
The tight confinement of the electric field across a mechanically-compliant narrow-gap capacitor brings the system into the quantum strong coupling regime under a red-sideband pump field. 
We demonstrate strong coupling between two mechanical modes, which is induced by two-tone parametric drives and mediated by a virtual photon in the cavity. 
The tunable inter-mechanical-mode coupling can be used to generate entanglement between the mechanical modes.  
\end{abstract}

\maketitle

Recently, application of optical techniques, such as laser cooling and trapping, to the control of a massive object has gained much interest.  
They have been promoting the progress in the control of the mechanical motions in a quantum regime, in conjunction with the technical advancement in the micro/nanofabrications as well as material engineering, which enable construction of massive mechanical oscillators with a high quality factor in varieties of designs and forms.
Cavity cooling of a mechanical oscillator, coupled to an electromagnetic resonators, down to the mechanical ground states was demonstrated through the radiation pressure in both microwave and optical regimes \cite{Teufel2011b, Chan2011a, Peterson2015a}.  
Observation of radiation-pressure shot noise \cite{Teufel2015b, Regal2013}, quantum nondemolition measurement and squeezing of a mechanical mode \cite{Teufel2015a, Schwab2015}, and bi-directional coherent transduction between microwave and optical photons \cite{Regal2014, Cleland2013}, are reported using various types of mechanical oscillators.  
In the microwave domain, it is of great interest to pursue coupling of a mechanical oscillator with a superconducting qubit to enhance its utility as a hybrid quantum system.  
Coherent coupling of a qubit with a bulk-mode photon in a mechanical oscillator \cite{Cleland2010, Teufel2015c, Sillanpaa2013} as well as with a traveling phonon using a surface acoustic wave \cite{Delsing} are observed.  
For the quantum control of a mechanical oscillator mode before the quantum state collapses, it is important that the phonon-photon coupling strength reaches the so-called quantum strong coupling regime, where the coupling strength $G$ exceeds the quantum state diffusion rate $\Gamma_{\rm{quant}}=n_{\rm{th}}\Gamma_m$ due to the warm environment~\cite{Verhagen2012a}.
Here, $ n_{\rm{th}}=k_B T_{\rm bath}/\hbar\omega _m$ is the thermal phonon occupation number at the reservoir temperature $T_{\rm bath}$, $k_B$ is Boltzmann constant, and $\omega_m$ and $\Gamma_m$ are the eigenfrequency and the decay rate of the mechanical oscillator mode, respectively. 

In this letter, we report the development of an electromechanical system using silicon-nitride membrane and a 3D loop-gap cavity.
The microwave electric field confined in an extremely small volume in the mechanically-compliant lumped-element circuit enhances the radiation pressure and enables the system to reach the quantum strong coupling regime.
We also show tunable strong electromechanical coupling between two mechanical modes of a massive object~\cite{stamper2015}.
While many studies concentrate on strong coupling between an electromagnetic mode and a mechanical mode, 
there are only a few multi-mode electro-(opto-)mechanical systems reported~\cite{sillanpaa2012, harris2014}.
Among these reports, only the system using cold atoms have reached the strong coupling regime~\cite{stamper2016}. 
The coupling between multiple mechanical modes of the massive object can further enrich their utility in the quantum information processing and in the exploration of the fundamental physics via entanglement between mechanical modes~\cite{Marquardt2012}.

%%% Figure 1%%%%
\begin{figure}[htb]
   \includegraphics[width=8.5cm,bb=0 0 1192 1498]{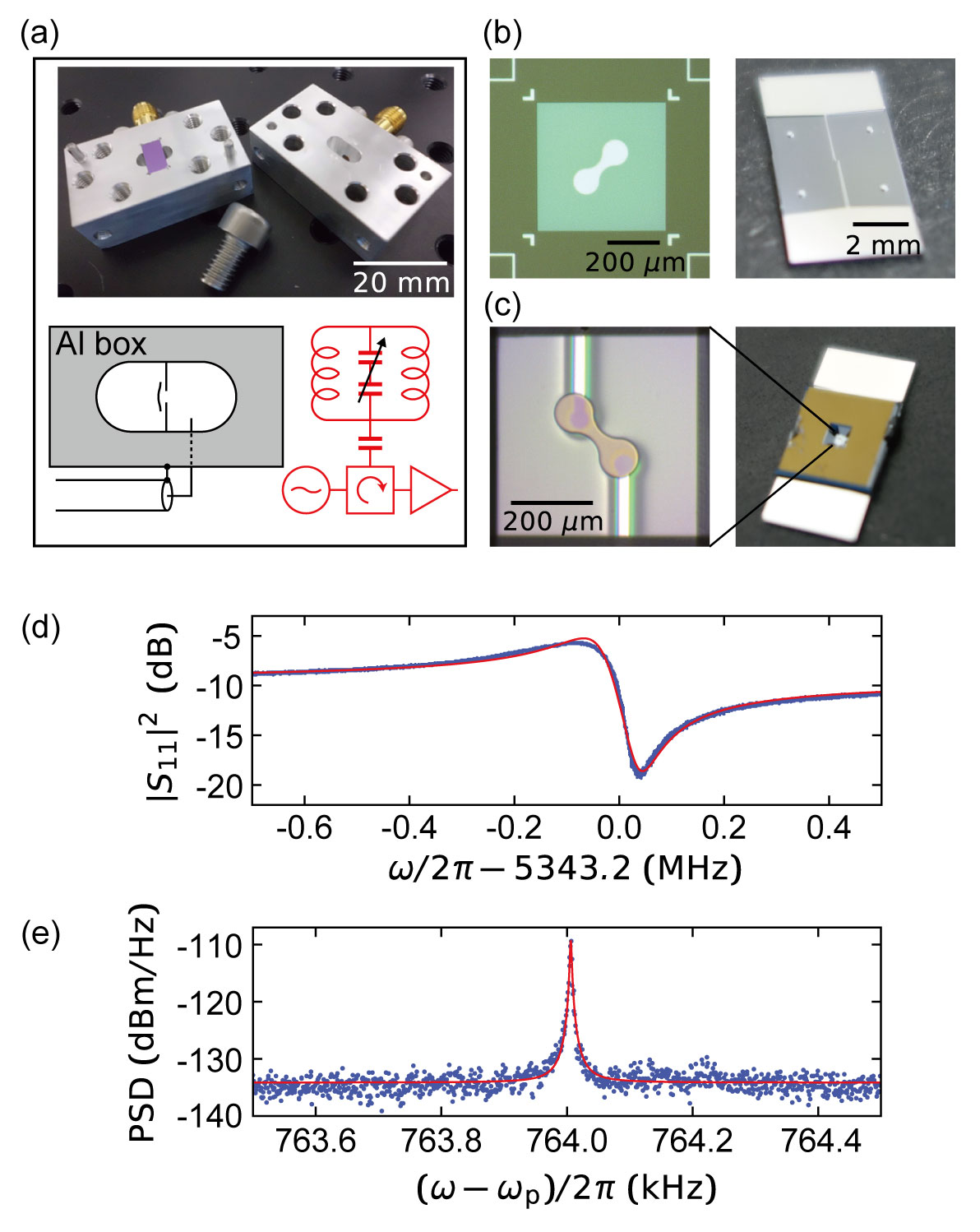}
\caption{
Physical implementation of the electromechanical system. (a) Three-dimensional loop-gap cavity constructed from an Al cavity and a capacitor circuit chip. External coupling with a coaxial signal line is adjusted by the protrusion length of the SMA connector pin into the cavity. 
Only one of the SMA connectors has a pin, forming a single-port cavity with a coupling constant $\kappa _{\rm ex}/2\pi = 30$ $\mathrm{kHz}$.
Lower panel shows a schematic of the measurement setup. 
Microwave reflection measurement of the electromechanical cavity is performed in a dilution refrigerator at around 10~mK.
(b) Photographs of a $500 \times 500$~$\mu$m Si$_3$N$_4$ membrane with an Al pad on top (left) and a bottom Si chip with Al electrodes (right).
The surface of the bottom chip is recessed by 100~$\mu$m, leaving the two electrodes and the four pillars high.
(c) Membrane chip flipped onto the bottom chip and glued with stycast on the edges.
(d) Microwave reflectance of the cavity, $|S_{11}|^2$ as a function of the probe frequency $\omega$ (blue). 
The data is taken at the probe power of $-150$~dBm corresponding to the average photon number in the cavity less than one.
The red line is a fit including the Fano effect due to the reflection from other microwave components.
(e) Thermal motion of the membrane observed in the noise spectrum of the cavity output. 
The power spectral density (PSD) is plotted as a function of the probe detuning relative to the pump frequency $\omega _{p}$ (blue). The red curve is a Lorentzian fit.
The cavity is pumped at the phonon red-sideband transition at a power of $P_{p} = -90$~dBm.
}
\label{fig1}
\end{figure}
%%%%%%%%%%%%

Our electromechanical system consists of a 3D aluminum cavity and a capacitor-circuit chip as shown in Fig.~1(a).
The capacitor circuit is constructed with an Al pad on a Si$_3$N$_4$ membrane chip and two electrodes on a bottom Si chip [Fig.~1(b)]. The 50-nm-thick membrane is made of stoichiometric Si$_3$N$_4$ and supported with a Si frame. 
A pair of parallel-plate capacitors are formed by placing the membrane chip on the bottom chip [Fig.~1(c)].
The gap between the pad on the membrane and the electrodes on the bottom chip is approximately 300 $\mathrm{nm}$, defined by the height of four pillars fabricated on the bottom chip.
Further details of the circuit fabrication process are described in supplementary material \cite{supp}.
The capacitor-circuit chip is loaded inside the 3D cavity to form a loop-gap cavity coupled with the mechanical motion of the membrane.
The electrodes on the chip are galvanically connected to the cavity wall.
In this configuration, the electric field is concentrated in the narrow gap capacitor, giving rise to the strong electromechanical coupling.

The empty 3D cavity has the lowest-frequency mode at $16.7$ $\mathrm{ GHz}$, which is lowered to $\omega _c/2\pi = 5.343$ $\mathrm{ GHz}$ when the capacitor-circuit chip is loaded [Fig.~1(d)]. 
The oscillation of the membrane modulates the parallel-plate capacitances and thus the cavity resonance frequency.
The electromechanical coupling strength is inversely proportional to the parallel-plate gap distance.
From the capacitor geometry and the cavity frequency, we estimate the single-photon electromechanical coupling strength of $g_1/2\pi = 7$ $\mathrm{ Hz}$ for the fundamental mode.
The internal quality factor of the cavity is $18\, 000$ at a probe power corresponding to the single-photon level inside the cavity.
At higher probe power, the quality factor reaches as high as $180\, 000$, presumably due to the saturation effect of the two-level systems associated with the cavity \cite{Cleland2012}.
Figure~1(e) shows the noise spectrum of the mechanical sideband. 
The vibration frequency and mechanical decay rate of the fundamental mode is found to be $\omega _{\rm 1}/2\pi =764~\mathrm{kHz}$ and $\Gamma _{\rm 1}/2\pi = 1.0~\mathrm{Hz}$, respectively. The decay rate is obtained from the ring-down measurement.

The canonical Hamiltonian of the electromechanical system is written as
\begin{equation}
\hat{H}=\hbar\omega _c\hat{a}^\dagger\hat{a}+\hbar\omega _{\rm 1}\hat{b}_1 ^\dagger\hat{b}_1+\hbar g_1\hat{a}^\dagger\hat{a}(\hat{b}_1+\hat{b}_1 ^\dagger ),
\end{equation}
where $\omega _c$ is the cavity resonant frequency, $\hat{a} (\hat{a}^\dagger)$ is the annihilation(creation) operator of the microwave photon, and $\hat{b}_1 (\hat{b}_1 ^\dagger)$ is the annihilation(creation) operator of the phonon.
When the pump microwave frequency $\omega_p$ is tuned at the phonon red-sideband $(\omega _p=\omega _c-\omega _{\rm 1})$,
the linearized Hamiltonian under the rotating-wave approximation reads \cite{Kippenberg2014}
\begin{equation}
\hat{H}=\hbar\omega _c\hat{a}^\dagger\hat{a}+\hbar\omega _{\rm 1}\hat{b}_1 ^\dagger\hat{b}_1+\hbar G_1(\alpha ) (\hat{a}\hat{b}_1 ^\dagger + \hat{a}^\dagger\hat{b}_1)
\end{equation}
where $G_1(\alpha )=g_1\sqrt{\alpha}$ is the parametrically-enhanced coupling where $\alpha$ is the microwave amplitude in the cavity.

%%% Figure 2%%%%
\begin{figure}[h]
   \includegraphics[width=8.5cm,bb=0 0 1210 1134]{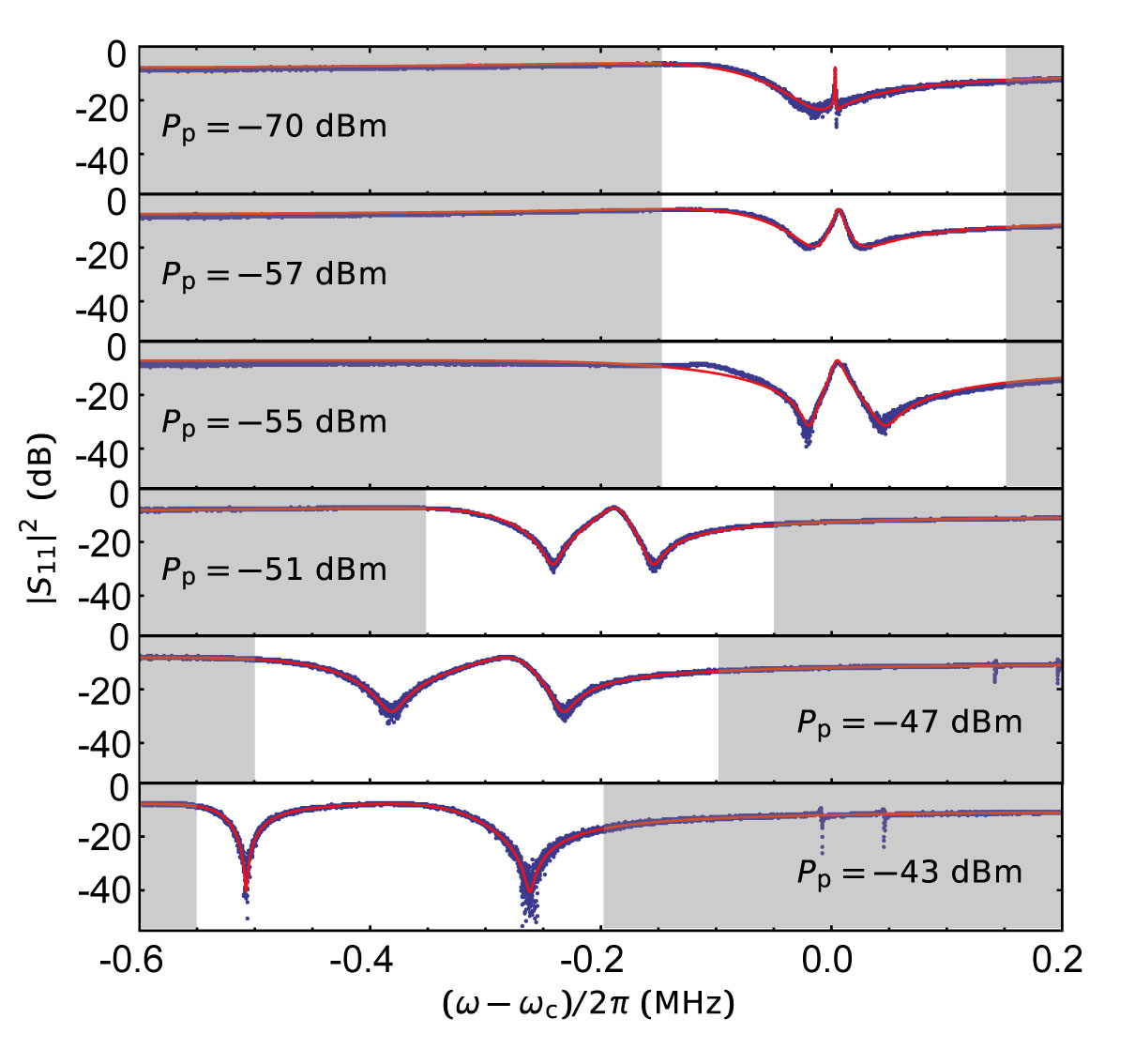}
\caption{
Probe reflectance $|S_{11}|^2$ for various pump powers $P_p$ (blue). The horizontal axis shows the probe frequency $\omega$ relative to the cavity resonance  frequency $\omega_c$ in the weak power limit. 
A narrow window generated by the electromechanically  induced transparency widens at higher pump power and exceeds the cavity linewidth.
The curves are the fit based on Eq.~(S1) in the Supplemental Material \cite{supp}.
The white zones are guides to the eyes.
}
\label{fig2}
\end{figure}
%%%%%%%%%%%%

We perform microwave spectroscopy of the cavity in the presence of the red-sideband pump field.
The reflection spectra $(|S_{11}|^2)$ of the probe field for various pump powers $P_p$ are shown in Fig.~2.
At the lowest pump power, the spectrum presents a feature known as electromechanically induced transparency \cite{Kippenberg2010}, showing a narrow transparent window within the broad cavity resonance. 
As the pump power is increased, the width of the window increases.
When the parametrically-enhanced coupling strength exceeds the cavity linewidth, the normal mode splitting is observed.
Other sharp dips observed at high pump powers are due to the weak electromechanical coupling with higher-frequency membrane modes.
In addition, we observe an intricate cavity frequency shift, which we presume originating from the nonlinearity of the cavity itself.

%%% Figure 3%%%%
\begin{figure}[b]
   \includegraphics[width=8.5cm,bb=0 0 1226 650]{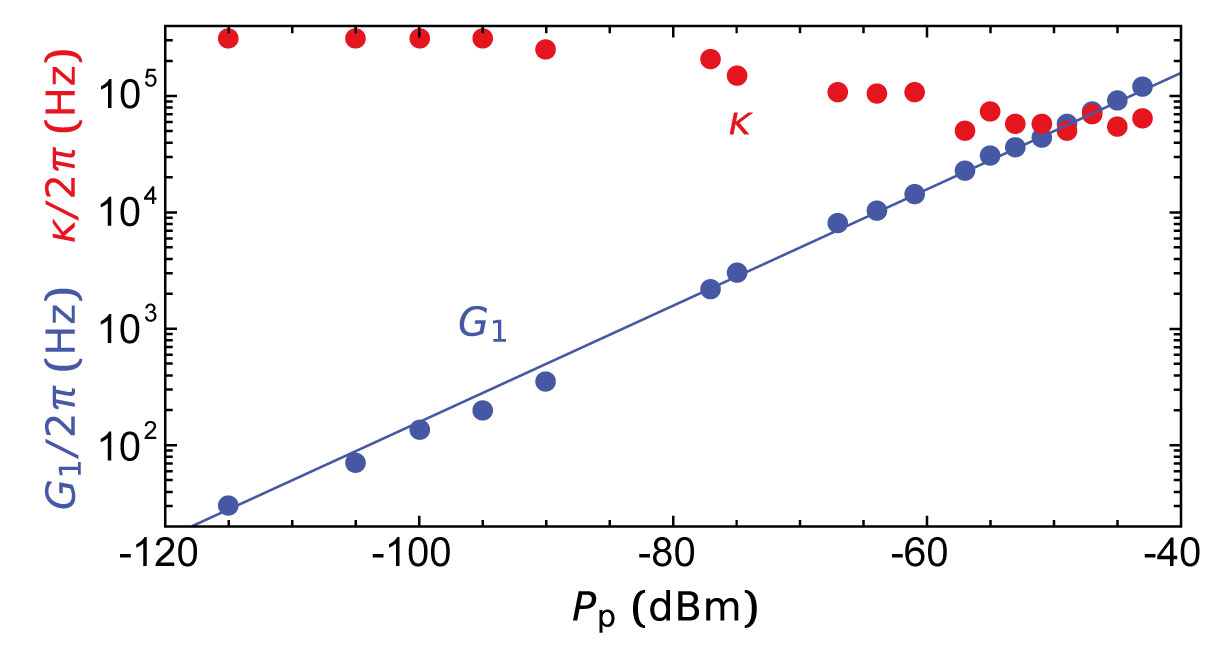}
\caption{
Electromechanical coupling $G_1$ (blue dots) and the cavity linewidth $\kappa$ (red dots) as a function of the pump power $P_p$.
The error bars are smaller than the dot size. The blue line indicates the square-root dependence on the pump power. 
}
\label{fig3}
\end{figure}
%%%%%%%%%%%%

Figure 3 shows the pump-power dependence of the electromechanical coupling $G_1$ and the cavity linewidth $\kappa$.
The cavity linewidth weakly depends on the pump power and is the narrowest at the highest pump power.
The electromechanical coupling increases proportionally to the square root of the drive power as expected.
A figure of merit of such hybrid quantum systems is often evaluated with the cooperativity $C=4G_1^2/\Gamma_1 \kappa$ \cite{Yuan2015a}.
However, when the bath temperature is much higher than the single-phonon temperature $T_{\rm ph} = \hbar \omega_{\rm 1} / k_B $, it is more appropriate to evaluate the system with the quantum cooperativity $C_{\rm{quant}}$ which expresses the strength of the electromechanical coupling with respect to the diffusion rate due to the thermal environment.
Quantum controllability of the electromechanical system requires the condition $C_{\rm{quant}} > 1$.
The maximum quantum cooperativity obtained for our system is
\begin{equation}
C_{\mathrm{quant}}=\frac{4G_1^2}{n_{\mathrm{th1}} \Gamma _{\rm 1} \kappa}=1750 \gg 1,
\end{equation}
where $n_{\rm th1}$ ($\approx 550$) is the thermal population of the membrane mode \cite{supp}.
Moreover, the electromechanical coupling strength is larger than the cavity linewidth and the mechanical decoherence rate in the thermal environment, meeting the condition for the strong coupling, $G_1>\kappa, n_{\mathrm{th}}\Gamma _{\rm 1}$.

%%% Figure 4%%%%
\begin{figure}[h]
   \includegraphics[width=8.5cm,bb=0 0 1236 1254]{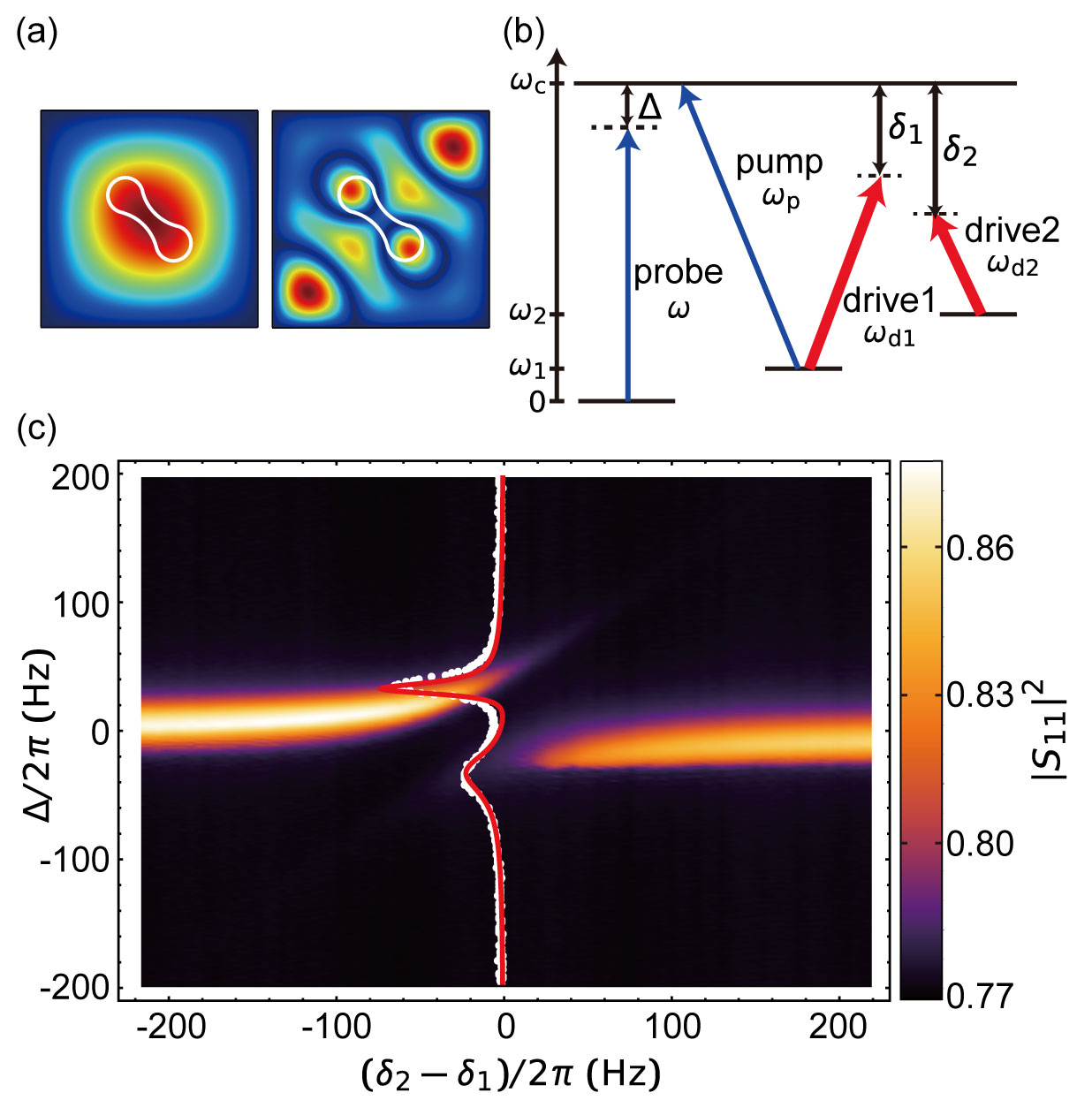}
\caption{
Strong coupling in multimode electromechanics. (a)~Calculated spatial mode profiles of the membrane with an Al pad evaporated on top: the fundamental mode
(mode~1; left) and a higher-frequency mode (mode~2; right). 
The white lines outline the pad shape.
(b) Schematic level diagram of the measurement of the parametrically induced multimode coupling. 
The coupling is induced by the two drive fields (drive 1 and 2). 
We monitor the resulting frequency shift through the electromechanically induced transparency using the pump and probe fields.
(c) Probe reflectance $|S_{11}|^2$ as a function of the drive detuning $\delta_2 - \delta_1$ and the probe detuning $\Delta = \omega - \omega_c$.
The white dots show the cross section showing the anticrossing at $\delta _1=\delta _2$. The red curve is the numerical simulation.
}
\label{fig4}
\end{figure}
%%%%%%%%%%%%

The particular design of the Al pad on the membrane also allows strong electromechanical coupling of the cavity with other membrane modes.
Figure 4(a) shows examples of simulated spatial modes of the membrane oscillations.
The left panel shows the profile of the fundamental mode (mode 1) with $\omega_{\rm 1}/2\pi = 764$ $\mathrm{kHz}$, where the mode structure is little affected by the presence of the pad.
On the other hand, the right panel shows the mode shape of a higher-frequency mode (mode 2) with the eigenfrequency of 
$\omega_{\rm 2} /2\pi = 2460 ~\mathrm{kHz}$.
The Al pad breaks the symmetry of the mode and largely modifies the mode shape.
Both of the mechanical modes have significant spatial overlap with the Al pad, resulting in strong coupling with the cavity mode.
The decay rate of mode 2 is found to be less than $\Gamma _2/2\pi <$ $1$ $\mathrm{Hz}$ from a measurement using the electromechanically induced transparency \cite{supp}. The measurement is limited by the bandwidth of the network analyzer.
The high quality factor ($>2\times 10^6$) of mode~2 can be explained from the spatial profile at the edge of the membrane:
the alternating signs of the mode function decouple the mode from the frame modes \cite{Parpia2011}.

These two mechanical modes can be coupled simultaneously to the microwave cavity mode with two drive fields.
Figure~4(b) is an energy diagram illustrating the mechanical and the resonator modes as well as the microwave tones used for the interrogation of the intermechanical coupling. 
We drive the cavity with two tones at the frequencies $\omega _{\rm d1}$ and $\omega _{\rm d2}$, both off-resonant from the microwave cavity frequency $\omega_c$.
Let us define the detunings $\delta_1 \equiv \omega_c - \omega_{1} - \omega_{\rm d1}$ and $\delta_2 \equiv \omega_c - \omega_{2} - \omega_{\rm d2}$. 
When the detunings are large enough, satisfying $\delta _1, \delta _2 >\kappa$, the microwave cavity mode can be adiabatically eliminated, leaving the two membrane modes parametrically coupled.
At two-photon resonance $\delta _1=\delta _2$, the interaction Hamiltonian between the two mechanical modes reads \cite{stamper2015}
\begin{equation}
\hat{H}_I=\eta (\hat{b}_{\rm 1}^\dagger \hat{b}_{\rm 2}  + \hat{b}_{\rm 1}\hat{b}_{\rm 2}^\dagger ),
\end{equation}
where $\hat{b}_{\rm 2}$ is the annihilation operator of mode~2, $\eta = G_1G_2 /\delta _1$ is the coupling strength between mode~1 and 2, where $G_1$ and $G_2$ are the parametrically-induced coupling strengths between the cavity mode and mode~1 and 2, respectively.

To observe the signature of the parametrically induced multimode coupling, we probe the shift of the fundamental mode via mechanically induced transparency, while sweeping the detuning of one of the drive fields [Fig.~4(b)]. 
Figure 4(c) shows the probe reflectance $|S_{11}|^2$ as a function of the drive detuning $\delta_2 - \delta_1$ and the probe detuning $\Delta \equiv \omega - \omega_c $.
The detuning $\delta _1/2\pi$ is fixed at $1.2$ $\mathrm{MHz}$.
The hybridization of the two mechanical modes are observed at around $\delta_1 - \delta_2 =0$, as an anticrossing in the spectrum.
The maximum coupling strength obtained between the two mechanical modes is $\eta /2\pi = 40$ $\mathrm{Hz}$ \cite{supp}.
A lower bound of the cooperativity for the phonon-phonon interaction is calculated as
\begin{equation}
C=\frac{4\eta ^2}{\Gamma _1 \Gamma _2 }>6400.
\end{equation}
The obtained coupling is already in the strong coupling regime. 

In the cross section shown in Fig.~4(c), an asymmetry between the hybridized modes is observed.
The asymmetry arises from the decoupling of one of the hybridized modes (upper peak in the figure) from the two drive fields:
The three strong fields including the pump field also act as cooling fields for the mechanical modes, resulting in the wider linewidths, while the upper hybridized mode is a dark mode for the two drive fields and is subject to less cooling.
The white curve in Fig. ~4(c) is the numerical simulation with parameters of $G_{1}/2\pi =12~\mathrm{kHz}$, $G_{2}/2\pi =3.7~\mathrm{kHz}$, $\delta_1/2\pi =1200~\mathrm{kHz}$, $G_p/2\pi =0.6~\mathrm {kHz}$ and $\kappa/2\pi = 200~\mathrm{kHz}$, where $G_p$ is the electromechanical coupling by the pump filed.

In conclusion, we demonstrated strong electromechanical coupling between a 3D loop-gap microwave cavity and a membrane mechanical oscillator.
The quatum cooperativity of the electromechanical coupling is found to be more than a thousand under a strong drive field.
As our system is constructed with a 3D microwave cavity, it can easily incorporate, e.g., a superconducting qubit, for realizing hybrid quantum systems capable for non-classical state manipulations and measurements.
We also implemented a multimode quantum electromechanical system by parametrically coupling two mechanical modes via the cavity mode with two drive microwave fields.
Utilizing multiple mechanical modes in the quantum regime can be an extra resource for quantum electromechanical systems. 
Multiple mechanical modes become a hybrid open quantum systems by the coupling the microwave field as an environment.
Possible applications include a long-lifetime quantum memory as well as devices for phonon-based quantum information processing.

This work was partly supported by the Project for Developing Innovation System of the Ministry of Education,
Culture, Sports, Science and Technology, Japan Society for the Promotion of Science KAKENHI (grant no. 26220601, 15H05461), 
and National Institute of Information and Communications Technology (NICT).

\bibliography{bibliography.bib}

\begin{thebibliography}{26}
\expandafter\ifx\csname natexlab\endcsname\relax\def\natexlab#1{#1}\fi
\expandafter\ifx\csname bibnamefont\endcsname\relax
  \def\bibnamefont#1{#1}\fi
\expandafter\ifx\csname bibfnamefont\endcsname\relax
  \def\bibfnamefont#1{#1}\fi
\expandafter\ifx\csname citenamefont\endcsname\relax
  \def\citenamefont#1{#1}\fi
\expandafter\ifx\csname url\endcsname\relax
  \def\url#1{\texttt{#1}}\fi
\expandafter\ifx\csname urlprefix\endcsname\relax\def\urlprefix{URL }\fi
\providecommand{\bibinfo}[2]{#2}
\providecommand{\eprint}[2][]{\url{#2}}

\bibitem[{\citenamefont{Teufel et~al.}(2011)\citenamefont{Teufel, Donner, Li,
  Harlow, Allman, Cicak, Sirois, Whittaker, Lehnert, and
  Simmonds}}]{Teufel2011b}
\bibinfo{author}{\bibfnamefont{J.~D.} \bibnamefont{Teufel}},
  \bibinfo{author}{\bibfnamefont{T.}~\bibnamefont{Donner}},
  \bibinfo{author}{\bibfnamefont{D.}~\bibnamefont{Li}},
  \bibinfo{author}{\bibfnamefont{J.~W.} \bibnamefont{Harlow}},
  \bibinfo{author}{\bibfnamefont{M.~S.} \bibnamefont{Allman}},
  \bibinfo{author}{\bibfnamefont{K.}~\bibnamefont{Cicak}},
  \bibinfo{author}{\bibfnamefont{A.~J.} \bibnamefont{Sirois}},
  \bibinfo{author}{\bibfnamefont{J.~D.} \bibnamefont{Whittaker}},
  \bibinfo{author}{\bibfnamefont{K.~W.} \bibnamefont{Lehnert}},
  \bibnamefont{and} \bibinfo{author}{\bibfnamefont{R.~W.}
  \bibnamefont{Simmonds}}, \bibinfo{journal}{Nature}
  \textbf{\bibinfo{volume}{475}}, \bibinfo{pages}{359} (\bibinfo{year}{2011}).

\bibitem[{\citenamefont{Chan et~al.}(2011)\citenamefont{Chan, Alegre,
  Safavi-Naeini, Hill, Krause, Gr\"{o}blacher, Aspelmeyer, and
  Painter}}]{Chan2011a}
\bibinfo{author}{\bibfnamefont{J.}~\bibnamefont{Chan}},
  \bibinfo{author}{\bibfnamefont{T.~P.~M.} \bibnamefont{Alegre}},
  \bibinfo{author}{\bibfnamefont{A.~H.} \bibnamefont{Safavi-Naeini}},
  \bibinfo{author}{\bibfnamefont{J.~T.} \bibnamefont{Hill}},
  \bibinfo{author}{\bibfnamefont{A.}~\bibnamefont{Krause}},
  \bibinfo{author}{\bibfnamefont{S.}~\bibnamefont{Gr\"{o}blacher}},
  \bibinfo{author}{\bibfnamefont{M.}~\bibnamefont{Aspelmeyer}},
  \bibnamefont{and} \bibinfo{author}{\bibfnamefont{O.}~\bibnamefont{Painter}},
  \bibinfo{journal}{Nature} \textbf{\bibinfo{volume}{478}}, \bibinfo{pages}{89}
  (\bibinfo{year}{2011}).

\bibitem[{\citenamefont{Peterson et~al.}(2015)\citenamefont{Peterson, Purdy,
  Kampel, Andrews, Yu, Lehnert, and Regal}}]{Peterson2015a}
\bibinfo{author}{\bibfnamefont{R.~W.} \bibnamefont{Peterson}},
  \bibinfo{author}{\bibfnamefont{T.~P.} \bibnamefont{Purdy}},
  \bibinfo{author}{\bibfnamefont{N.~S.} \bibnamefont{Kampel}},
  \bibinfo{author}{\bibfnamefont{R.~W.} \bibnamefont{Andrews}},
  \bibinfo{author}{\bibfnamefont{P.~L.} \bibnamefont{Yu}},
  \bibinfo{author}{\bibfnamefont{K.~W.} \bibnamefont{Lehnert}},
  \bibnamefont{and} \bibinfo{author}{\bibfnamefont{C.~A.} \bibnamefont{Regal}},
  \bibinfo{journal}{arXiv:1510.03911}  (\bibinfo{year}{2015}).

\bibitem[{\citenamefont{Teufel et~al.}(2016)\citenamefont{Teufel, Lecocq, and
  Simmonds}}]{Teufel2015b}
\bibinfo{author}{\bibfnamefont{J.~D.} \bibnamefont{Teufel}},
  \bibinfo{author}{\bibfnamefont{F.}~\bibnamefont{Lecocq}}, \bibnamefont{and}
  \bibinfo{author}{\bibfnamefont{R.~W.} \bibnamefont{Simmonds}},
  \bibinfo{journal}{Phys. Rev. Lett.} \textbf{\bibinfo{volume}{116}},
  \bibinfo{pages}{013602} (\bibinfo{year}{2016}).

\bibitem[{\citenamefont{Purdy et~al.}(2013)\citenamefont{Purdy, Peterson, and
  Regal}}]{Regal2013}
\bibinfo{author}{\bibfnamefont{T.~P.} \bibnamefont{Purdy}},
  \bibinfo{author}{\bibfnamefont{R.~W.} \bibnamefont{Peterson}},
  \bibnamefont{and} \bibinfo{author}{\bibfnamefont{C.~A.} \bibnamefont{Regal}},
  \bibinfo{journal}{Science} \textbf{\bibinfo{volume}{339}},
  \bibinfo{pages}{801} (\bibinfo{year}{2013}).

\bibitem[{\citenamefont{Lecocq et~al.}(2015{\natexlab{a}})\citenamefont{Lecocq,
  Clark, Simmonds, Aumentado, and Teufel}}]{Teufel2015a}
\bibinfo{author}{\bibfnamefont{F.}~\bibnamefont{Lecocq}},
  \bibinfo{author}{\bibfnamefont{J.~B.} \bibnamefont{Clark}},
  \bibinfo{author}{\bibfnamefont{R.~W.} \bibnamefont{Simmonds}},
  \bibinfo{author}{\bibfnamefont{J.}~\bibnamefont{Aumentado}},
  \bibnamefont{and} \bibinfo{author}{\bibfnamefont{J.~D.}
  \bibnamefont{Teufel}}, \bibinfo{journal}{Phys. Rev. X}
  \textbf{\bibinfo{volume}{5}}, \bibinfo{pages}{041037}
  (\bibinfo{year}{2015}{\natexlab{a}}).

\bibitem[{\citenamefont{Wollman et~al.}(2015)\citenamefont{Wollman, Lei,
  Weinstein, J.~Suh, Marquardt, Clerk, and Schwab}}]{Schwab2015}
\bibinfo{author}{\bibfnamefont{E.~E.} \bibnamefont{Wollman}},
  \bibinfo{author}{\bibfnamefont{C.~U.} \bibnamefont{Lei}},
  \bibinfo{author}{\bibfnamefont{A.~J.} \bibnamefont{Weinstein}},
  \bibinfo{author}{\bibfnamefont{A.~K.} \bibnamefont{J.~Suh}},
  \bibinfo{author}{\bibfnamefont{F.}~\bibnamefont{Marquardt}},
  \bibinfo{author}{\bibfnamefont{A.~A.} \bibnamefont{Clerk}}, \bibnamefont{and}
  \bibinfo{author}{\bibfnamefont{K.~C.} \bibnamefont{Schwab}},
  \bibinfo{journal}{Science} \textbf{\bibinfo{volume}{349}},
  \bibinfo{pages}{952} (\bibinfo{year}{2015}).

\bibitem[{\citenamefont{Andrews et~al.}(2014)\citenamefont{Andrews, Peterson,
  Purdy, Cicak, Simmonds, Regal, and Lehnert}}]{Regal2014}
\bibinfo{author}{\bibfnamefont{R.}~\bibnamefont{Andrews}},
  \bibinfo{author}{\bibfnamefont{R.}~\bibnamefont{Peterson}},
  \bibinfo{author}{\bibfnamefont{T.~P.} \bibnamefont{Purdy}},
  \bibinfo{author}{\bibfnamefont{K.}~\bibnamefont{Cicak}},
  \bibinfo{author}{\bibfnamefont{R.}~\bibnamefont{Simmonds}},
  \bibinfo{author}{\bibfnamefont{C.~A.} \bibnamefont{Regal}}, \bibnamefont{and}
  \bibinfo{author}{\bibfnamefont{K.~W.} \bibnamefont{Lehnert}},
  \bibinfo{journal}{Nat. Phys.} \textbf{\bibinfo{volume}{10}},
  \bibinfo{pages}{321} (\bibinfo{year}{2014}).

\bibitem[{\citenamefont{Bochmann et~al.}(2013)\citenamefont{Bochmann,
  Vainsencher, Awschalom, and Cleland}}]{Cleland2013}
\bibinfo{author}{\bibfnamefont{J.}~\bibnamefont{Bochmann}},
  \bibinfo{author}{\bibfnamefont{A.}~\bibnamefont{Vainsencher}},
  \bibinfo{author}{\bibfnamefont{D.~D.} \bibnamefont{Awschalom}},
  \bibnamefont{and} \bibinfo{author}{\bibfnamefont{A.~N.}
  \bibnamefont{Cleland}}, \bibinfo{journal}{Nat. Phys.}
  \textbf{\bibinfo{volume}{9}}, \bibinfo{pages}{712} (\bibinfo{year}{2013}).

\bibitem[{\citenamefont{O'Connell et~al.}(2010)\citenamefont{O'Connell,
  Hofheinz, Ansmann, Bialczak, Lenander, Lucero, and Neeley}}]{Cleland2010}
\bibinfo{author}{\bibfnamefont{A.~D.} \bibnamefont{O'Connell}},
  \bibinfo{author}{\bibfnamefont{M.}~\bibnamefont{Hofheinz}},
  \bibinfo{author}{\bibfnamefont{M.}~\bibnamefont{Ansmann}},
  \bibinfo{author}{\bibfnamefont{R.~C.} \bibnamefont{Bialczak}},
  \bibinfo{author}{\bibfnamefont{M.}~\bibnamefont{Lenander}},
  \bibinfo{author}{\bibfnamefont{E.}~\bibnamefont{Lucero}}, \bibnamefont{and}
  \bibinfo{author}{\bibfnamefont{M.}~\bibnamefont{Neeley}},
  \bibinfo{journal}{Nature} \textbf{\bibinfo{volume}{464}},
  \bibinfo{pages}{697} (\bibinfo{year}{2010}).

\bibitem[{\citenamefont{Lecocq et~al.}(2015{\natexlab{b}})\citenamefont{Lecocq,
  Teufel, Aumentado, and Simmonds}}]{Teufel2015c}
\bibinfo{author}{\bibfnamefont{F.}~\bibnamefont{Lecocq}},
  \bibinfo{author}{\bibfnamefont{J.~D.} \bibnamefont{Teufel}},
  \bibinfo{author}{\bibfnamefont{J.}~\bibnamefont{Aumentado}},
  \bibnamefont{and} \bibinfo{author}{\bibfnamefont{R.~W.}
  \bibnamefont{Simmonds}}, \bibinfo{journal}{Nat. Phys.}
  \textbf{\bibinfo{volume}{11}}, \bibinfo{pages}{635}
  (\bibinfo{year}{2015}{\natexlab{b}}).

\bibitem[{\citenamefont{Pirkkalainen et~al.}(2013)\citenamefont{Pirkkalainen,
  Cho, Li, Paraoanu, Hakonen, and Sillanp\"{a}\"{a}}}]{Sillanpaa2013}
\bibinfo{author}{\bibfnamefont{J.-M.} \bibnamefont{Pirkkalainen}},
  \bibinfo{author}{\bibfnamefont{S.~U.} \bibnamefont{Cho}},
  \bibinfo{author}{\bibfnamefont{J.}~\bibnamefont{Li}},
  \bibinfo{author}{\bibfnamefont{G.~S.} \bibnamefont{Paraoanu}},
  \bibinfo{author}{\bibfnamefont{P.~J.} \bibnamefont{Hakonen}},
  \bibnamefont{and} \bibinfo{author}{\bibfnamefont{M.~A.}
  \bibnamefont{Sillanp\"{a}\"{a}}}, \bibinfo{journal}{Nature}
  \textbf{\bibinfo{volume}{494}}, \bibinfo{pages}{211} (\bibinfo{year}{2013}).

\bibitem[{\citenamefont{Gustafsson et~al.}(2014)\citenamefont{Gustafsson, Aref,
  Kockum, Ekstrom, Johansson, and Delsing}}]{Delsing}
\bibinfo{author}{\bibfnamefont{M.~V.} \bibnamefont{Gustafsson}},
  \bibinfo{author}{\bibfnamefont{T.}~\bibnamefont{Aref}},
  \bibinfo{author}{\bibfnamefont{A.~F.} \bibnamefont{Kockum}},
  \bibinfo{author}{\bibfnamefont{M.~K.} \bibnamefont{Ekstrom}},
  \bibinfo{author}{\bibfnamefont{G.}~\bibnamefont{Johansson}},
  \bibnamefont{and} \bibinfo{author}{\bibfnamefont{P.}~\bibnamefont{Delsing}},
  \bibinfo{journal}{Science} \textbf{\bibinfo{volume}{346}},
  \bibinfo{pages}{207} (\bibinfo{year}{2014}).

\bibitem[{\citenamefont{Verhagen et~al.}(2012)\citenamefont{Verhagen,
  Del\'{e}glise, Weis, Schliesser, and Kippenberg}}]{Verhagen2012a}
\bibinfo{author}{\bibfnamefont{E.}~\bibnamefont{Verhagen}},
  \bibinfo{author}{\bibfnamefont{S.}~\bibnamefont{Del\'{e}glise}},
  \bibinfo{author}{\bibfnamefont{S.}~\bibnamefont{Weis}},
  \bibinfo{author}{\bibfnamefont{A.}~\bibnamefont{Schliesser}},
  \bibnamefont{and} \bibinfo{author}{\bibfnamefont{T.~J.}
  \bibnamefont{Kippenberg}}, \bibinfo{journal}{Nature}
  \textbf{\bibinfo{volume}{482}}, \bibinfo{pages}{63} (\bibinfo{year}{2012}).

\bibitem[{\citenamefont{Buchmann and Stamper-Kurn}(2015)}]{stamper2015}
\bibinfo{author}{\bibfnamefont{L.~F.} \bibnamefont{Buchmann}} \bibnamefont{and}
  \bibinfo{author}{\bibfnamefont{D.~M.} \bibnamefont{Stamper-Kurn}},
  \bibinfo{journal}{Phys. Rev. A} \textbf{\bibinfo{volume}{92}},
  \bibinfo{pages}{013851} (\bibinfo{year}{2015}).

\bibitem[{\citenamefont{Massel et~al.}(2012)\citenamefont{Massel, Cho,
  Pirkkalainen, Hakonen, Heikkila, and Sillanpaa}}]{sillanpaa2012}
\bibinfo{author}{\bibfnamefont{F.}~\bibnamefont{Massel}},
  \bibinfo{author}{\bibfnamefont{S.~U.} \bibnamefont{Cho}},
  \bibinfo{author}{\bibfnamefont{J.-M.} \bibnamefont{Pirkkalainen}},
  \bibinfo{author}{\bibfnamefont{P.~J.} \bibnamefont{Hakonen}},
  \bibinfo{author}{\bibfnamefont{T.~T.} \bibnamefont{Heikkila}},
  \bibnamefont{and} \bibinfo{author}{\bibfnamefont{M.~A.}
  \bibnamefont{Sillanpaa}}, \bibinfo{journal}{Nat. Commun.}
  \textbf{\bibinfo{volume}{3}}, \bibinfo{pages}{987} (\bibinfo{year}{2012}).

\bibitem[{\citenamefont{Shkarin et~al.}(2014)\citenamefont{Shkarin,
  Flowers-Jacobs, Hoch, Kashkanova, Deutsch, Reichel, and Harris}}]{harris2014}
\bibinfo{author}{\bibfnamefont{A.~B.} \bibnamefont{Shkarin}},
  \bibinfo{author}{\bibfnamefont{N.~E.} \bibnamefont{Flowers-Jacobs}},
  \bibinfo{author}{\bibfnamefont{S.~W.} \bibnamefont{Hoch}},
  \bibinfo{author}{\bibfnamefont{A.~D.} \bibnamefont{Kashkanova}},
  \bibinfo{author}{\bibfnamefont{C.}~\bibnamefont{Deutsch}},
  \bibinfo{author}{\bibfnamefont{J.}~\bibnamefont{Reichel}}, \bibnamefont{and}
  \bibinfo{author}{\bibfnamefont{J.~G.~E.} \bibnamefont{Harris}},
  \bibinfo{journal}{Phys. Rev. Lett.} \textbf{\bibinfo{volume}{112}},
  \bibinfo{pages}{013602} (\bibinfo{year}{2014}).

\bibitem[{\citenamefont{Spethmann et~al.}(2016)\citenamefont{Spethmann, Kohler,
  Schreppler, Buchmann, and Stamper-Kurn}}]{stamper2016}
\bibinfo{author}{\bibfnamefont{N.}~\bibnamefont{Spethmann}},
  \bibinfo{author}{\bibfnamefont{J.}~\bibnamefont{Kohler}},
  \bibinfo{author}{\bibfnamefont{S.}~\bibnamefont{Schreppler}},
  \bibinfo{author}{\bibfnamefont{L.}~\bibnamefont{Buchmann}}, \bibnamefont{and}
  \bibinfo{author}{\bibfnamefont{D.~M.} \bibnamefont{Stamper-Kurn}},
  \bibinfo{journal}{Nat. Phys.} \textbf{\bibinfo{volume}{12}},
  \bibinfo{pages}{27} (\bibinfo{year}{2016}).

\bibitem[{\citenamefont{Schmidt et~al.}(2012)\citenamefont{Schmidt, Ludwig, and
  Marquardt}}]{Marquardt2012}
\bibinfo{author}{\bibfnamefont{M.}~\bibnamefont{Schmidt}},
  \bibinfo{author}{\bibfnamefont{M.}~\bibnamefont{Ludwig}}, \bibnamefont{and}
  \bibinfo{author}{\bibfnamefont{F.}~\bibnamefont{Marquardt}},
  \bibinfo{journal}{New J. Phys.} \textbf{\bibinfo{volume}{14}},
  \bibinfo{pages}{125005} (\bibinfo{year}{2012}).

\bibitem[{sup(2016)}]{supp}
\emph{\bibinfo{title}{Supplemental Material}} (\bibinfo{year}{2016}).

\bibitem[{\citenamefont{Megrant et~al.}(2012)\citenamefont{Megrant, Neill,
  Barends, Chiaro, Chen, Feigl, Kelly, Lucero, Mariantoni, O'Malley
  et~al.}}]{Cleland2012}
\bibinfo{author}{\bibfnamefont{A.}~\bibnamefont{Megrant}},
  \bibinfo{author}{\bibfnamefont{C.}~\bibnamefont{Neill}},
  \bibinfo{author}{\bibfnamefont{R.}~\bibnamefont{Barends}},
  \bibinfo{author}{\bibfnamefont{B.}~\bibnamefont{Chiaro}},
  \bibinfo{author}{\bibfnamefont{Y.}~\bibnamefont{Chen}},
  \bibinfo{author}{\bibfnamefont{L.}~\bibnamefont{Feigl}},
  \bibinfo{author}{\bibfnamefont{J.}~\bibnamefont{Kelly}},
  \bibinfo{author}{\bibfnamefont{E.}~\bibnamefont{Lucero}},
  \bibinfo{author}{\bibfnamefont{M.}~\bibnamefont{Mariantoni}},
  \bibinfo{author}{\bibfnamefont{P.~J.~J.} \bibnamefont{O'Malley}},
  \bibnamefont{et~al.}, \bibinfo{journal}{Appl. Phys. Lett.}
  \textbf{\bibinfo{volume}{100}}, \bibinfo{pages}{113510}
  (\bibinfo{year}{2012}).

\bibitem[{\citenamefont{M.~Aspelmeyer}(2014)}]{Kippenberg2014}
\bibinfo{author}{\bibfnamefont{F.~M.} \bibnamefont{M.~Aspelmeyer},
  \bibfnamefont{T.~J.~Kippenberg}}, \bibinfo{journal}{Rev. Mod. Phys.}
  \textbf{\bibinfo{volume}{86}}, \bibinfo{pages}{1391} (\bibinfo{year}{2014}).

\bibitem[{\citenamefont{Weis et~al.}(2010)\citenamefont{Weis, Riviere,
  Del\'{e}glise, Gavartin, Arcizet, Schliesser, and
  Kippenberg}}]{Kippenberg2010}
\bibinfo{author}{\bibfnamefont{S.}~\bibnamefont{Weis}},
  \bibinfo{author}{\bibfnamefont{R.}~\bibnamefont{Riviere}},
  \bibinfo{author}{\bibfnamefont{S.}~\bibnamefont{Del\'{e}glise}},
  \bibinfo{author}{\bibfnamefont{E.}~\bibnamefont{Gavartin}},
  \bibinfo{author}{\bibfnamefont{O.}~\bibnamefont{Arcizet}},
  \bibinfo{author}{\bibfnamefont{A.}~\bibnamefont{Schliesser}},
  \bibnamefont{and} \bibinfo{author}{\bibfnamefont{T.~J.}
  \bibnamefont{Kippenberg}}, \bibinfo{journal}{Science}
  \textbf{\bibinfo{volume}{330}}, \bibinfo{pages}{1520} (\bibinfo{year}{2010}).

\bibitem[{\citenamefont{Yuan et~al.}(2015)\citenamefont{Yuan, Singh, Blanter,
  and Steele}}]{Yuan2015a}
\bibinfo{author}{\bibfnamefont{M.}~\bibnamefont{Yuan}},
  \bibinfo{author}{\bibfnamefont{V.}~\bibnamefont{Singh}},
  \bibinfo{author}{\bibfnamefont{Y.~M.} \bibnamefont{Blanter}},
  \bibnamefont{and} \bibinfo{author}{\bibfnamefont{G.~A.}
  \bibnamefont{Steele}}, \bibinfo{journal}{Nat. Commun.}
  \textbf{\bibinfo{volume}{6}}, \bibinfo{pages}{8491} (\bibinfo{year}{2015}).

\bibitem[{\citenamefont{Wilson-Rae et~al.}(2011)\citenamefont{Wilson-Rae,
  Barton, Verbridge, Southworth, Ilic, Craighead, and Parpia}}]{Parpia2011}
\bibinfo{author}{\bibfnamefont{I.}~\bibnamefont{Wilson-Rae}},
  \bibinfo{author}{\bibfnamefont{R.~A.} \bibnamefont{Barton}},
  \bibinfo{author}{\bibfnamefont{S.~S.} \bibnamefont{Verbridge}},
  \bibinfo{author}{\bibfnamefont{D.~R.} \bibnamefont{Southworth}},
  \bibinfo{author}{\bibfnamefont{B.}~\bibnamefont{Ilic}},
  \bibinfo{author}{\bibfnamefont{H.~G.} \bibnamefont{Craighead}},
  \bibnamefont{and} \bibinfo{author}{\bibfnamefont{J.~M.}
  \bibnamefont{Parpia}}, \bibinfo{journal}{Phys. Rev. Lett.}
  \textbf{\bibinfo{volume}{106}}, \bibinfo{pages}{047205}
  (\bibinfo{year}{2011}).

\bibitem[{\citenamefont{Zwickl et~al.}(2008)\citenamefont{Zwickl, Shanks,
  Jayich, Yang, Jayich, Thompson, and Harris}}]{Harris2008}
\bibinfo{author}{\bibfnamefont{B.~M.} \bibnamefont{Zwickl}},
  \bibinfo{author}{\bibfnamefont{W.~E.} \bibnamefont{Shanks}},
  \bibinfo{author}{\bibfnamefont{A.~M.} \bibnamefont{Jayich}},
  \bibinfo{author}{\bibfnamefont{C.}~\bibnamefont{Yang}},
  \bibinfo{author}{\bibfnamefont{A.~C.~B.} \bibnamefont{Jayich}},
  \bibinfo{author}{\bibfnamefont{J.~D.} \bibnamefont{Thompson}},
  \bibnamefont{and} \bibinfo{author}{\bibfnamefont{J.~G.~E.}
  \bibnamefont{Harris}}, \bibinfo{journal}{Appl. Phys. Lett.}
  \textbf{\bibinfo{volume}{92}}, \bibinfo{pages}{103125}
  (\bibinfo{year}{2008}).

\end{thebibliography}

\section*{Supplemental Material}
\subsection{Sample Fabrication}

\renewcommand{\thefigure}{S\arabic{figure}}
\renewcommand{\thetable}{S\arabic{table}}
\renewcommand{\theequation}{S\arabic{equation}}
\setcounter{figure}{0}
\setcounter{equation}{0}
%%% Figure A%%%%
\begin{figure}
   \includegraphics[width=8.5cm,bb=0 0 1184 956]{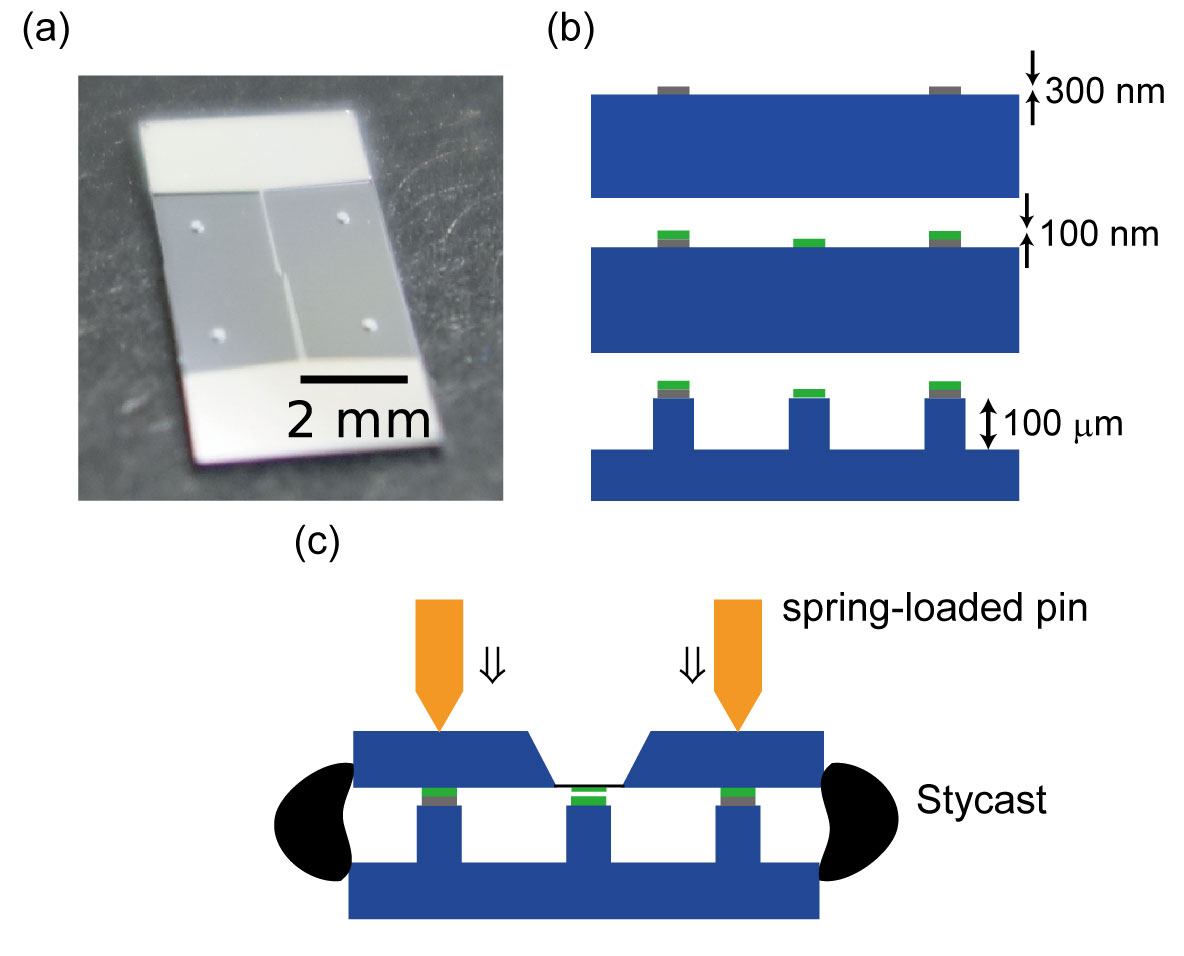}
\caption{
Sample fabrication method. (a) Photograph of the bottom Si chip containing two Al electrodes and four support pillars.
(b) Fabrication process. After evaporation and wet-etching of two layers of Al (brown and green), the Si substrate is recessed with Deep RIE.
(c) Schematic of the flip-chip mounting and gluing procedure.  
}
\label{fig1sp}
\end{figure}
%%%%%%%%%%%%

Figure \ref{fig1sp}(a) and (b) shows the bottom Si chip and its fabrication process.
All the aluminum circuits and structures are made by Al films evaporated and patterned with photo-lithography with S1805 photo-resist followed by wet-etching with TMAH etchant.  
First, 300-nm-tall Al pillars are formed on a Si substrate ($\rho > 10$~k$\Omega\cdot$cm).  
Then, 100-nm-thick electrodes are evaporated and patterned with photo-lithography and wet-etching.
The Si substrate surrounding the circuit is recessed by a deep reactive ion etching (Deep RIE).
This design prevents small dust particles from getting caught between the membrane and the bottom chip.
The depth of the etching is approximately 100 $\mathrm{\mu m}$.

The Si$_3$N$_4$ membrane is purchased from Norcada Inc.~\cite{Harris2008}. It is made of stoichiometric high-tension Si$_3$N$_4$ and spans on a Si frame with a square window of $500 \times 500$~$\mu$m. 
A 30-nm-thick Al pad is formed on the 50-nm-thick membrane.

Figure \ref{fig1sp}(c) shows the flip-chip placement and adhesion procedure.
We clean the surface of the bottom chip thoroughly and flip the membrane chip onto the bottom chip.
The membrane chip is pressed against the pillars with spring-loaded pins and is glued with epoxy (Stycast 1285).
As shown in Fig.~1(c) in the main article, the interference effect between the reflected light from the top and the bottom electrodes varies the apparent color of the electrode at the overlapping region.
The color depends on the distance between the electrodes, which we use for a qualitative check of the gap formation.

%%% Figure A'%%%%
\begin{figure}[h]
   \includegraphics[width=8.5cm,bb=0 0 984 1190]{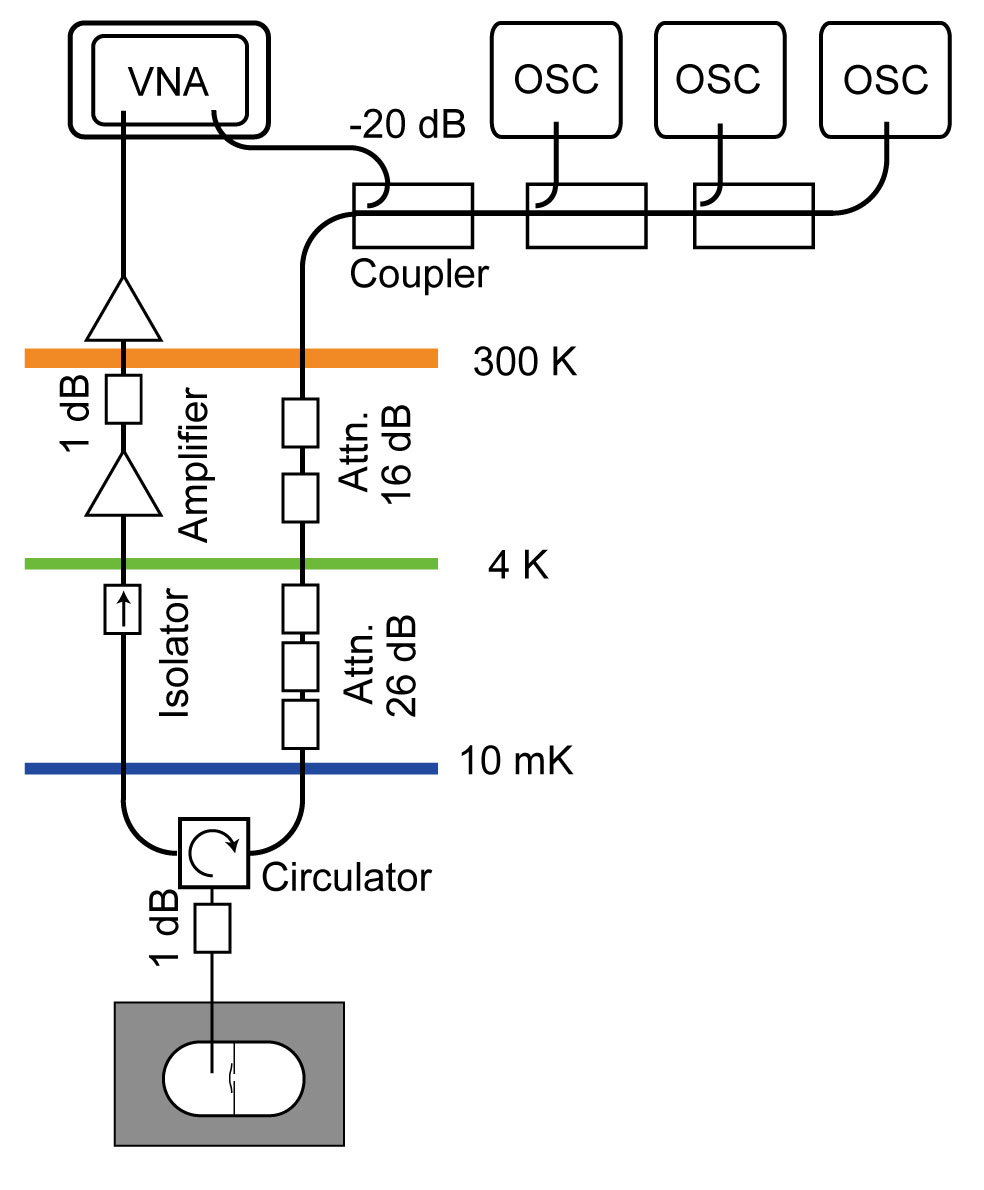}
\caption{
Measurement apparatus.
}
\label{fig6sp}
\end{figure}
%%%%%%%%%%%%

\subsection{Measurement setup}
Figure~\ref{fig6sp} illustrates the microwave measurement setup.
The loop-gap cavity is cooled to 10~mK in a dilution refrigerator.
Attenuators are mounted on the input coaxial cable at each plate of the  fridge to prevent the thermal noise from entering the cavity.  
The total attenuation through the input line is 50.3 dB at 6~GHz.
For the output line, an isolator and a circulator protect the cavity from the thermal noise and the amplifier noise. The signal is amplified by the  two low-noise amplifiers at 4~K and room temperature.
We use a vector network analyzer to observe the reflection coefficient $S_{11}$ of the electromechanical system.
There are up to three microwave oscillators for the pump and the drives, which are combined through a series of directional couplers.

%% Figure C%%%%
\begin{figure}[h]
   \includegraphics[width=8.5cm,bb=0 0 1218 822]{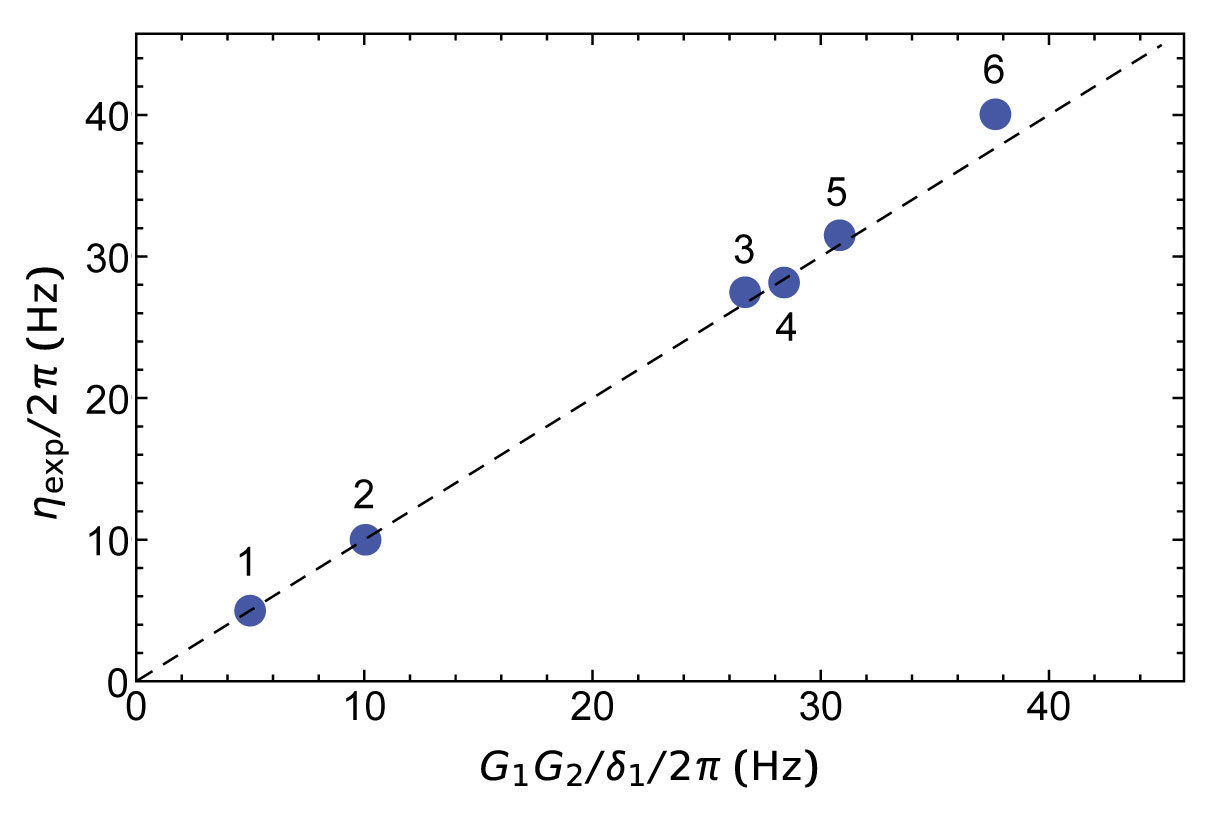}
\caption{
Strength of the parametrically-induced interaction between the two mechanical modes in the membrane.
We vary the drive powers and detuning $\delta _1$.
The horizontal axis shows the calculated mechanical coupling $G_1G_2/\delta _1$. 
The number on each dot indicates the corresponding parameter set in Table~S1. 
The vertical axis is the experimentally obtained coupling strength between the two modes.
The dashed line indicates $\eta =G_1G_2/\delta _1$.
}
\label{fig3sp}
\end{figure}
%%%%%%%%%%%

\begin{table}
\begin{tabular}{c|ccccc}\hline
\# & $G_{\rm 1}$ & $G_{\rm 2}$ & $\delta _1/2\pi$ & $\eta _{\rm cal}/2\pi$ & $\eta _{\rm exp}/2\pi$ \\
 &  $(\mathrm{kHz})$ & $(\mathrm{kHz})$ &(kHz) &(Hz)&(Hz)\\\hline \hline
1& 8.1 & 0.74 & 1200 & 5.0 & 5.0\\
2& 11.4 & 1.05 & 1200 & 10.0 & 9.8 \\
3& 11.4 & 2.80 & 1200 & 26.7 & 27.4\\
4& 8.1 & 1.05 & 300 & 28.4 & 28.6\\
5& 11.4 & 2.97 & 1100 & 30.8 & 31.6\\
6& 12.1 & 3.73 & 1200 & 37.7 & 40.0\\\hline
\end{tabular}
\caption{
Coupling strength.
We measure the coupling strength $\eta _{\exp}$ as a function of the drive powers and the detuning.
}
\label{table}
\end{table}

\subsection{Determination of coupling strength between mechanical modes}
When the two drive fields are on resonance $\delta_1-\delta_2=0$, the coupling strength between the two mechanical modes is given as $\eta _{\rm cal} = G_1G_2/\delta_1$, where $G_i$ $(i=1,2)$ is the electromechanical coupling of mode $i$ with the cavity mode and $\delta _i$ is the detuning of the drive field $i$.
We measure the $G_i$ by the electromechanically induced transparency of each drive power.
Table \ref{table} shows the compiled data of the measured coupling strength $\eta_{\rm exp}$ between the two mechanical modes for various sets of $G_{\rm 1}$, $G_{\rm 2}$, and $\delta_1$.  
Figure \ref{fig3sp} represents the coupling strength of these data.
From these values of $G_i$ and intra-cavity photon with input-output theory, 
we observe the single photon coupling strength $g_1/2\pi = 7.20~\mathrm{Hz}$ and $g_2/2\pi = 1.04~\mathrm{Hz}$.
The measured values of $\eta_{exp}$ agree well with the theoretical values.  

%%% Figure D%%%%
\begin{figure}[h]
   \includegraphics[width=8.5cm,bb=0 0 1232 718]{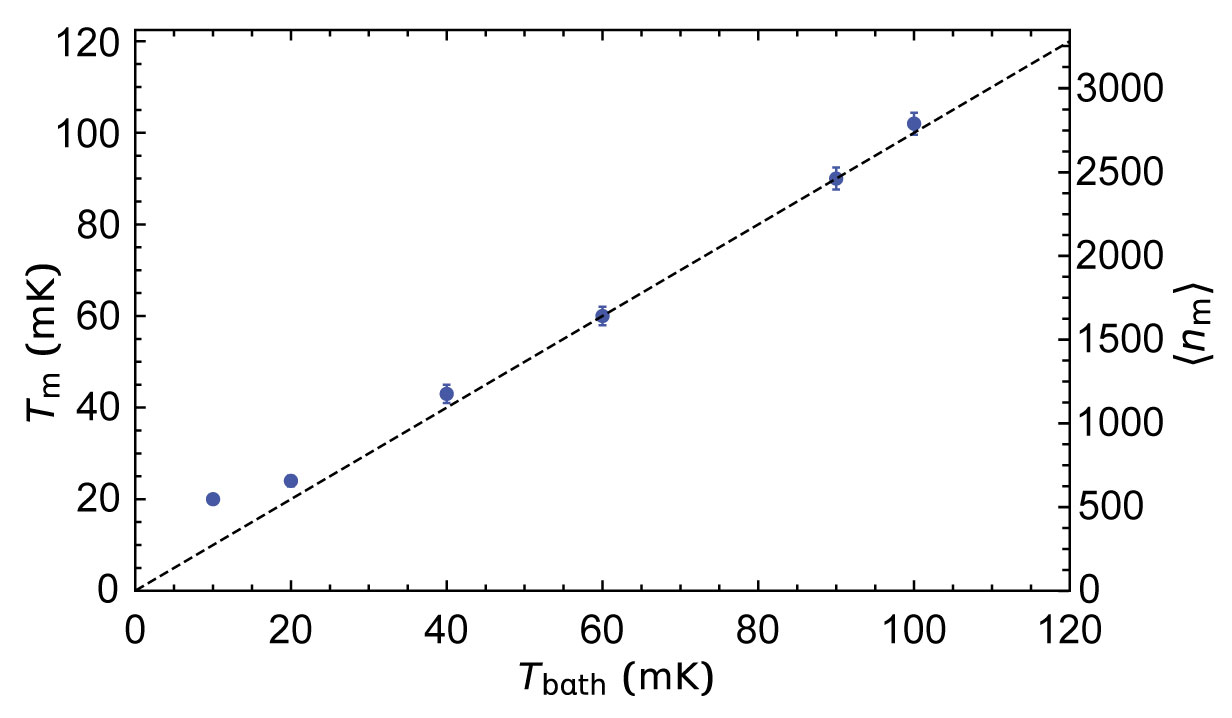}
\caption{
Temperature calibration. 
The horizontal axis is the temperature obtained from a $\mathrm{RuO}_2$ thermometer mounted on the mixing-chamber plate of the dilution refrigerator. 
The left (right) vertical axis is the effective temperature (the average phonon number) of the fundamental mode obtained from the peak area of the noise spectrum.
}
\label{fig4sp}
\end{figure}
%%%%%%%%%%%%

\subsection{Temperature calibration}
In order to obtain the phonon number in the fundamental mode, we measure the mechanical noise spectra at different dilution fridge temperature for the calibration.
A weak probe field is used to avoid cooling of the membrane by the probe field as well as the direct heating of the system from the input microwave.
We first plot the normalized area of the noise spectrum as a function of the bath temperature, and the data at higher temperatures to calibrate the phonon number with respect to the area of the spectra.
As shown in Fig.~\ref{fig4sp} the lowest phonon number down to 550 was observed without additional cooling.

%% Figure D%%%%
\begin{figure}[b]
   \includegraphics[width=8.5cm,bb=0 0 1310 688]{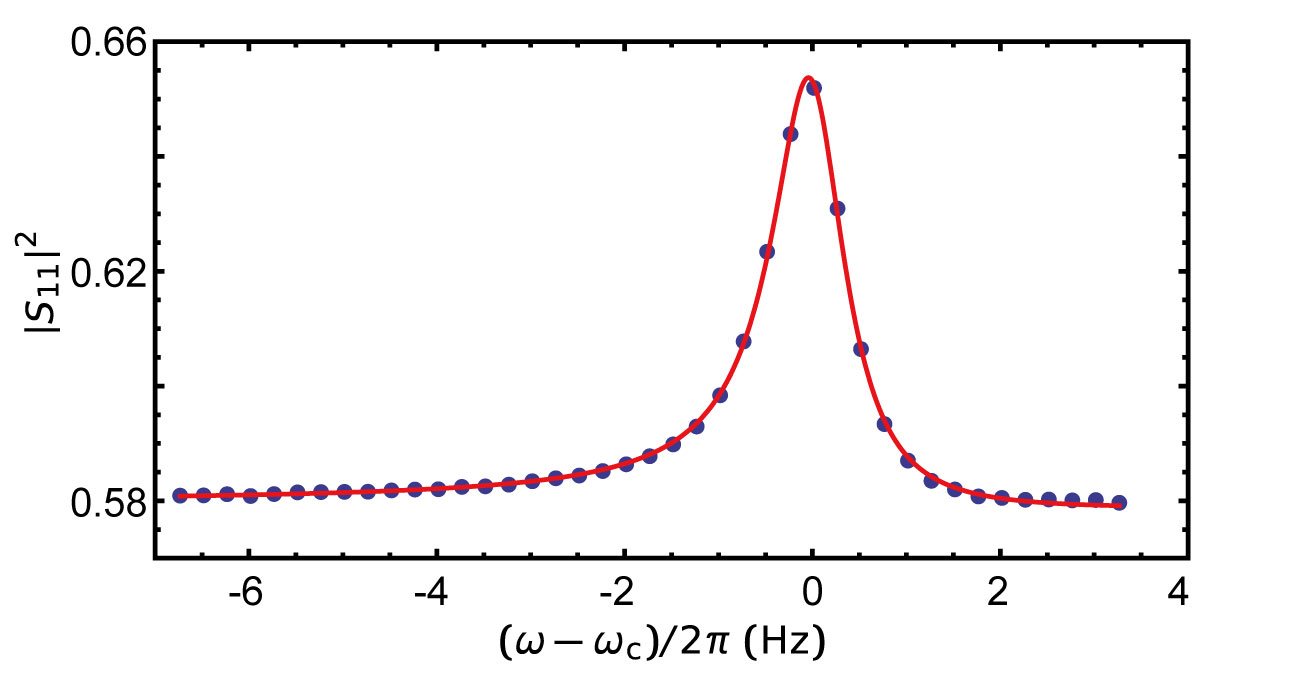}
\caption{
Linewidth measurement of the higher-frequency mode. 
Mechanically induced transparency due to the higher-frequency mode. The reflectivity of the probe field is measured under a weak pump field at $\omega_p = \omega_c - \omega_{2}$.
}
\label{fig5sp}
\end{figure}
%%%%%%%%%%%
\subsection{Decay rate measurement of the higher-frequency mechanical mode}
In order to obtain the intrinsic linewidth of the membrane mode, the pump power is decreased as much as possible.
Figure \ref{fig5sp} shows the mechanical induced transparency for the higher-frequency mode with the weak pump.
Here, the resolution of 1~Hz is limited by the bandwidth of the network analyzer.
The fitting curve is a similar function of Eqs. (S1) and (S2).

\subsection{Fitting function of the cavity reflection}
For fitting the cavity reflectivity spectrum in the presence of a mechanically induced transparency window, we use a function,
\begin{equation}
f(\omega )=\left| a_1 \exp (i\theta _1)+ a_0 \left[ 1-\sqrt{\xi \kappa} g(\omega )\right] \right| ^2,
\end{equation}
where
\begin{equation}
g(\omega )\equiv \frac{\sqrt{\xi \kappa}}{-i(\omega - \omega _c) +\kappa /2 +G_1^2/[-i(\omega - \Delta_{\rm p} )+\Gamma _{\rm 1}/2]},
\end{equation}
$a_1$ and $\theta _1$ is the amplitude and the phase of the stray microwave field caused by reflections at various microwave components along the measurement line, $\kappa $ is the total linewidth of the cavity, 
$\xi = \kappa _{\rm ex}/\kappa$ is the ratio of the external coupling to the total linewidth of the cavity, $\omega _c$ is the cavity resonant frequency, $\Delta_{\rm p} = \omega _c - \omega _p - \omega _{\rm m}$ is the detuning of the electromechanical coupling, $\Gamma _{\rm 1}$ is the linewidth of the fundamental mechanical mode, and $G_1$ is the electromechanical coupling.
The fitting parameters are $(a_1, \theta _1, \kappa, \eta, \omega _c, G_1)$, while fixing $\Gamma _{\rm 1}/2\pi$ at $1$~Hz, which we obtained from an independent measurement.
\end{document}